\documentclass[AMA]{article} 
\usepackage{anyfontsize} %

\usepackage{colortbl}   
\usepackage{xcolor}     

\usepackage{steinmetz}

\usepackage{siunitx}
\usepackage{subcaption}

\usepackage{circuitikz}
\usepackage{tikzscale}
\usepackage{multirow}
\usepackage{xspace}
\usepackage{authblk}

\usepackage[backend=bibtex, sorting=none]{biblatex}
\bibliography{references}

\usepackage[acronym,nomain,nonumberlist]{glossaries}
\usepackage{glossary-inline}     
\setacronymstyle{long-short} 
\makenoidxglossaries

\renewcommand*{\glossarysection}[2][]{\textbf{Abbreviations:} }%
\renewcommand*{\glsinlinedescformat}[3]{}%
\renewcommand*{\glsinlineemptydescformat}[2]{}%
\glsdisablehyper

\newacronym{rf}{RF}{radiofrequency}
\newacronym{snr}{SNR}{Signal-to-Noise Ratio}
\newacronym{em}{EM}{Electromagnetic}
\newacronym{sibc}{SIBC}{Surface Impedance Boundary Conditions}
\newacronym{pec}{PEC}{Perfect Electric Conductor}
\newacronym{ac}{AC}{Alternate Current}
\newacronym{dc}{DC}{Direct Current}
\newacronym{nrmse}{NRMSE}{Normalized Root Mean Square Errors}
\newacronym{pla}{PLA}{PolyLactic Acid}
\newacronym{pva}{PVA}{PolyVinyl Alcohol}
\newacronym{mrcods}{MRCODS}{MRI RF Coil Open Data Standard}
\newacronym{pcb}{PCB}{Printed Circuit Board}
\newacronym{gpio}{GPIO}{General Purpose Input/Output}
\newacronym{dsv}{DSV}{Diameter Spherical Volume}
\newacronym{emi}{EMI}{Electromagnetic Interference}
\newacronym{txrx}{TXRX}{Transmit-receive}
\newacronym{lna}{LNA}{Low Noise Amplifier}
\newacronym{fff}{FFF}{Fused Filament Fabrication}
\newacronym{mri}{MRI}{Magnetic Resonance Imaging}
\newacronym{rms}{RMS}{Root Mean Square}
\newacronym{sar}{SAR}{Specific Absorption Rate}
\newacronym{vna}{VNA}{Vector Network Analyzer}
\newacronym[longplural=Regions Of Interest]{roi}{ROI}{Region Of Interest}

\graphicspath{{./images/}{./figures/}}

\newcommand{\Bz}{B\textsubscript{0}\xspace}
\newcommand{\Bop}{B\textsubscript{1}\textsuperscript{+}\xspace}

\title{Birth of the Coil: another Milestone towards a fully reproducible low-field MRI scanner for head-imaging}

\author[1]{Umberto Zanovello}

\author[2]{Julia Pfitzer}

\author[3]{Ariane Ernst}

\author[4]{Joshua Harper}

\author[5]{Jack Hayhoe}

\author[6]{Tri Nguyen}

\author[3]{Lukas Winter}

\author[7,8]{Nicola De Zanche}

\affil[1]{Istituto Nazionale di Ricerca Metrologica (INRiM), Torino,Italy}

\affil[2]{Institute of Biomedical Imaging, Graz University of Technology, Graz, Austria}

\affil[3]{Physikalisch-Technische Bundesanstalt (PTB) Braunschweig and Berlin, Berlin, Germany}

\affil[4]{Department of Technology and Applied Sciences, Universidad Comunera, Asuncion, Paraguay}

\affil[5]{Mechanical Engineering, University of Alberta, Edmonton, Canada}

\affil[6]{Electrical and Computer Engineering, University of Alberta, Edmonton, Canada}

\affil[7]{OncologyUniversity of Alberta, Edmonton, Canada}

\affil[8]{Medical Physics, Cross Cancer Institute, Edmonton, Canada}

\begin{document}

\maketitle

\abstract{Low-field magnetic resonance imaging (MRI) provides an accessible, portable, and low-cost alternative to high-field scanners, expanding diagnostic imaging to point-of-care settings. However, widespread adoption is fundamentally hindered by a severely reduced signal-to-noise ratio (SNR). At low frequencies, radiofrequency (RF) coil conductor losses — rather than tissue sample losses — predominantly govern the system's total noise, making meticulous RF coil optimization critical to recovering image quality. This work presents an open-source, optimized solenoid head coil tailored for the \SI{50}{\milli\tesla} open-source scanner (OSI\textsuperscript{2} ONE v2.1). The paper validates production reproducibility across three independent international institutions and introduce an open-source connector with integrated digital circuitry for coil identification and DC or logic signals. Comprehensive benchtop measurements, Electromagnetic Interference (EMI) coupling analysis, Specific Absorption Rate (SAR) safety simulations, and phantom and human volunteer imaging confirm the design's efficacy, safety, and reproducibility. The results of the paper, when combined with the material provided in the open-source dedicated repositories, set the basis for a fully reliable and reproducible component for the open-source OSI\textsuperscript{2} ONE MRI scanner. In addition, the same optimization strategy and design material can be exploited for designing other RF coils for imaging of other body parts.}

\maketitle
\renewcommand\thefootnote{}
\footnotetext{\printnoidxglossary[type=\acronymtype,style=inline,sort=word]}%

\renewcommand\thefootnote{\fnsymbol{footnote}}
\setcounter{footnote}{1}

\section{Introduction}

Accessibility, portability, reduced costs, increased safety are just few of the reasons justifying the revitalized interest in low-field \gls{mri}\cite{bib:campbell1, bib:hennig1}. The major problem hindering the diffusion of low-field \gls{mri} systems is the reduced \gls{snr} which, in the low-field limit, is proportional to the 7/4th power of \Bz \cite{bib:edelstein1, bib:oreilly2}. Despite this limitation, low-field \gls{mri} systems are growing and complete open-source low-field \gls{mri} solutions are becoming available. One such example is set by OSI\textsuperscript{2}\cite{bib:osiiWebsite} with the OSI\textsuperscript{2} ONE \SI{50}{\milli\tesla} scanner\cite{bib:osii_one}. The open-source approach, when supported by a thorough inclusion of all the design and production file, as well as a comprehensive documentation, potentially ensure a rapid and effective advancement of the device, repairability of the components and, eventually, that these technologies reach patients\cite{bib:winter1}. This paper sits in this context by designing and characterizing an open-source \gls{rf} coil for head-imaging, specifically optimized for operation with the OSI\textsuperscript{2} ONE v2.1 scanner.

The reason of the 7/4th power proportionality between \gls{snr} and \Bz resides in the higher impact that \gls{rf} coil losses have on the signal noise with respect to sample losses and suggests the paramount importance of a proper design of the \gls{rf} coil in a low-field system. In high-field clinical system, birdcage coils are the most commonly used solution, providing high \gls{snr} for horizontal bore magnets generating a \Bz along their main axis. The magnet used by the OSI\textsuperscript{2} ONE scanner\cite{bib:osii_one} generates a \Bz along the vertical direction\cite{bib:osiiMagnet}. Whereas birdcage coils can still be used for vertical \Bz systems\cite{bib:giovannetti5}, their complexity makes them no more a practical solution. In contrast, solenoid coils becomes an optimum candidate, especially at very low frequencies, where they demonstrated better \gls{snr} when compared to other volume coils\cite{bib:marocco1, bib:blasiak1}. Therefore, the paper optimizes a solenoid structure showing how different choices affect the \gls{rf} coil performance. In doing this, the paper gives particular emphasis to the reproducibility of the device by comparing the performance of three independently characterized \gls{rf} coil replicas; one at the Istituto Nazionale di Ricerca Metrologica (INRiM, Italy), one at the Physikalisch-Technische Bundesanstalt (PTB, Germany) and one at the Graz University of Technology (TU Graz, Austria). 

While recognizing the foremost importance of the radiating element in the \gls{rf} coil system, the paper also focuses on the design and realization of an open-source connector, entirely made of off-the-shelf contacts and realized through rapid prototyping technologies\cite{bib:hayhoe1}. This aims addressing one limitation of current open-source \gls{mri} system, where the \gls{rf} coil connector is simply supposed to connect the \gls{rf} coil to the \gls{mri} system in order to provide \gls{rf} transmit power and receive signal. The proposed connector design also integrates an I\textsuperscript{2}C-EEPROM and digital circuitry for storing \gls{rf} coil information and providing \gls{dc} or logic signals when needed (e.g., on-coil temperature sensors, detuning of PIN diodes, etc.). The connector provides therefore the same functionality, at a fraction of the cost, of those typically found in mainstream commercial \gls{mri} systems and complies with the \gls{mrcods} open-standard for coil identification\cite{bib:mrcods}.

The paper is structured in the following sections. Sections \ref{sec:rfCoilDesign-Method} and \ref{sec:rfCoilDesign-Results} focus on the design and optimization of the \gls{rf} coil. Sections \ref{sec:benchtopMeasurements-Method} and \ref{sec:benchtopMeasurements-Results} deal with the measurement of the \gls{rf} coil performance in terms of Q factor, S11 bandwidth and insertion loss of the \gls{rf} connector. Sections \ref{sec:emiCoupling-Method} and \ref{sec:emiCoupling-Results} investigates the \gls{rf} coil performance in terms of \gls{emi} coupling showing a possible solution to improve its common-mode rejection ratio. Sections \ref{sec:safetyAssesment-Method} and  \ref{sec:safetyAssesment-Results} analyze, through numerical simulations, the \gls{sar} generated by the coil on a human body model. Sections \ref{sec:rfConnector-Method} and \ref{sec:rfConnector-Results} describe the \gls{rf} coil connector. Finally, sections \ref{sec:mriExperiments-Method} and \ref{sec:mriExperiments-Results} demonstrate the performance of the coil when used for imaging of phantom and human volunteer.

\section{Methods}
\subsection{RF Coil Design}
\label{sec:rfCoilDesign-Method}
\begin{figure}
  \centering
  \includegraphics[width=0.8\textwidth]{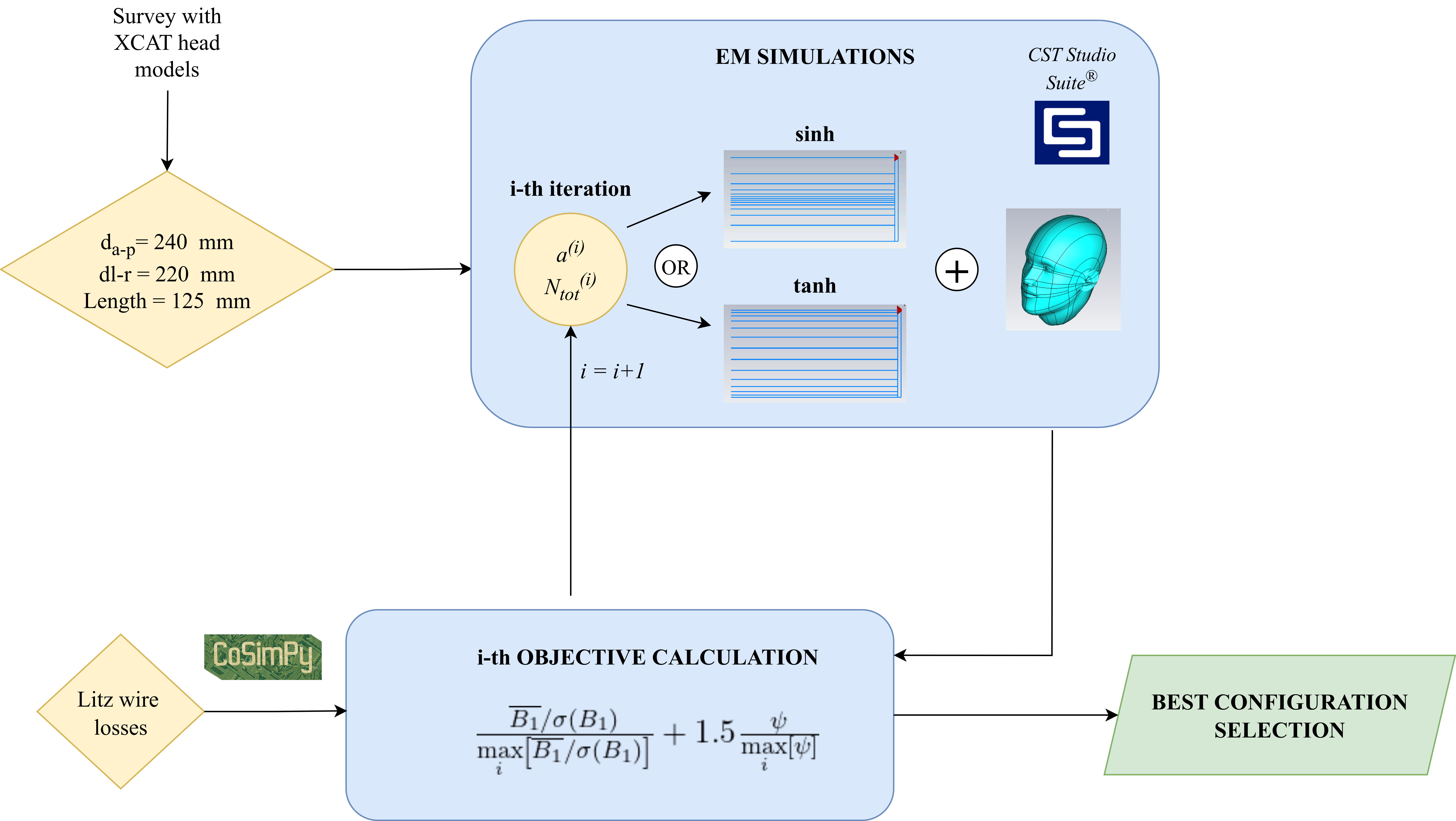}
  \caption{Workflow followed for the \gls{rf} coil design and optimization. Refer to the main text for the meaning of the symbols}
  \label{fig:rfcoilOptimizationWorkflow}
\end{figure}
\figurename\,\ref{fig:rfcoilOptimizationWorkflow} shows the \gls{rf} coil design and optimization workflow, which begins by specifying the \gls{rf} coil length and diameter. Although a larger coil is desirable to fit most human heads and reduce claustrophobia, a narrower coil improves the filling factor and increases the distance to the \gls{rf} shield thus improving the \gls{snr}\cite{bib:gruber1, bib:parsa1}. To reduce claustrophobia, the \gls{rf} coil length was limited from the nose to the top of the head. To maximize the filling factor, the \gls{rf} coil adopted an elliptical axial profile and the diameters were selected on an average head based on a survey of seven human body anatomies (4 males and 3 females) from the XCAT library of anatomical models\cite{bib:segars1}. The \gls{rf} coil length was limited to \SI{125}{\milli\meter} and different anterior-posterior ($d_{a-p}$) and left-right ($d_{l-r}$) diameters were used and equaled \SI{240}{\milli\meter} and \SI{220}{\milli\meter}, respectively.

These dimensions represented the input for a series of full-wave \gls{em} simulations aimed at identifying the optimum coil configuration in terms of number and distribution of windings along the main axis of the coil . Equations \eqref{eq:sinhDistr}, \eqref{eq:tanhDistr} parametrized the winding positions, $z$, along the coil axis for the two different configurations investigated in the optimization process.
\begin{subequations}
  \begin{align}
    z_s(n) = \frac{L/2}{\sinh(\pi/a)}\sinh\left(\frac{n\Delta-\pi}{a}\right)\label{eq:sinhDistr}\\
    z_t(n) = \frac{L/2}{\tanh(\pi/a)}\tanh\left(\frac{n\Delta-\pi}{a}\right)\label{eq:tanhDistr}
  \end{align}
\end{subequations}
where $L$ is the length of the coil, $a$ adjusts the maximum and minimum distance between adjacent windings along the coil axis, $n$ is a discrete parameter ranging from 0 to $N_{tot}-1$, with $N_{tot}$ being the number of coil windings, and $\Delta$ is equal to $\frac{2\pi}{N_{tot}-1}$.
In \eqref{eq:sinhDistr} the conductors are more gathered in the center of the solenoid and in \eqref{eq:tanhDistr} on its edges. In the case of an infinite value of $a$, both configurations reduce to a standard constant-pitch solenoid, with equal distance between neighboring windings. 


The x- and y-coordinates of each winding were determined according to the following relationship:
\begin{equation}\label{eq:xyCoords}
    \frac{x^2}{(d_{a-p}/2)^2} + \frac{y^2}{(d_{l-r}/2)^2} = 1
\end{equation}
where $t$ is the parametrization variable ranging from $-\pi$ to $\pi$. 

The coil optimization was performed by simulating both the sinh and tanh conductor distributions with different values of the $a$ parameter and total number of windings, $N_{tot}$. Preliminary investigations showed that there were no improvements on the \gls{rf} coil performance for values of $a$ higher than \num{4}, setting its upper limit during the optimization process. The minimum value of $a$ depended on the $N_{tot}$ parameter in the specific simulation. In particular, for a given $N_{tot}$, the minimum value of $a$ was selected to guarantee a maximum ratio between the conductor diameter (selected a priori equal to \SI{3}{\milli\meter}) and the minimum conductor axial distance of \num{0.7}. Limiting the maximum value of such a ratio was justified by its impact on the proximity effects, which negatively affect the \gls{rf} coil losses and performances\cite{bib:medhurst1,bib:blasiak1,bib:butterworth1}. According to theoretical calculations\cite{bib:medhurst1, bib:butterworth1}, limiting this value to \num{0.7} corresponds to a ratio of the coil resistance to the resistance of a straight wire with the same length lower than \num{3.3}.

For each $N_{tot}$, ranging from \numrange{8}{14}, simulations accounted for three logarithmically spaced values of the $a$ parameter and the following objective was computed:
\begin{equation}
  \text{objective}=\underbrace{\frac{\overline{B_1}/\sigma(B_1)}{\max\limits_i \left[\overline{B_1}/\sigma(B_1)\right]}}_{A} + 1.5\underbrace{\frac{\psi}{\max\limits_i [\psi]}}_{B}
  \label{eq:optObjective}
\end{equation}
where the overline represents the spatial average, $\sigma$ represents the spatial standard deviation, B\textsubscript{1} is the main \gls{rf} magnetic field component, $\psi$ is the \gls{rf} coil sensitivity computed as $\frac{\overline{B_1}}{2\sqrt{P_{acc}}}$, with $P_{acc}$ being the accepted power into the \gls{rf} coil, and $\max\limits_i$ means the maximum among all simulations. The objective is therefore composed of two contributions, the first (A) reflects on the \Bop homogeneity and the second (B) on the average \gls{snr} and the amount of power needed to flip the magnetization. Since the \gls{snr} is critical at low-field, the second contribution is slightly favored over the first by a factor of \num{1.5}.

The configuration which maximized equation \eqref{eq:optObjective} was identified in terms of $a$ and $N_{tot}$. Nine additional \gls{em} simulations refined the optimum $a$ value for the resulted optimum $N_{tot}$ value.

\gls{em} simulations were performed with the frequency domain solver of CST Studio Suite\cite{bib:cst} from \SIrange{1.7}{2.7}{\mega\hertz} enabling the adaptive mesh refinement. Results were extracted at \SI{2.1}{\mega\hertz} which corresponds to the Larmor frequency of the OSI\textsuperscript{2} \SI{50}{\milli\tesla} magnet\cite{bib:osiiMagnet}. Simulation included a cylindrical shield with a \SI{280}{\milli\meter} diameter and \SI{400}{\milli\meter} length. The shield was simulated adopting the \gls{sibc} instead of the more common \gls{pec} approximation. By modeling the skin effect, \gls{sibc} guarantee an accurate estimation of the Joule losses\cite{bib:fawzi,bib:yuferev1,bib:zilberti1} which, for low-field \gls{rf} coil, can be comparable or even higher than sample losses\cite{bib:giovannetti1, bib:brown1, bib:blasiak1}. An obvious consequence of the fact that coil losses are not strongly sample-dominated is that conductor losses must be thoroughly accounted for. For the coil widings, special attention must be paid to skin and proximity effects\cite{bib:terman1}. Both of them increase the conductor losses and the adoption of a proper Litz wire proved to be an effective solution to address the issue for low-field \gls{rf} coil design\cite{bib:grafendorfer1, bib:stormont1, bib:giovannetti2}.
Parameters like the number and diameter of the strands, the twist angle and the wire diameter noticeably affect the wire performance in terms of losses, and the optimum solution depends on the specific application\cite{bib:butterworth1,bib:grafendorfer1}. Unfortunately, it is not straightforward to reliably simulate such a wire since simulations would require very thin mesh elements to discretize each strand making the simulation process unsustainable. Butterworth\cite{bib:butterworth1} provided an analytical expression for the \gls{ac} resistance per unit length of a single layer stranded wire solenoid, $R_{ac}$:
\begin{equation}
  \begin{split}
  R_{ac} = R_{dc}\sec\alpha\left\{1+F(z)+s^2d^2G(z)\left[\frac{2}{D^2}\left(1-\frac{1}{2}\sin^2\alpha\right) \right.\right.\\
  \left.\left.+\frac{u_n}{\Delta_s^2}\left( 1-\frac{1}{4}\sin^2\alpha\right)\right] \right\}
  \label{eq:litzTheoretical}
  \end{split}
\end{equation}
where:
\begin{itemize}
  \item[] $R_{dc}$ is the \gls{dc} resistance per unit length of the equivalent solid wire;
  \item[] $s$ is the number of strands;
  \item[] $d$ is the diameter of the single strand in meters;
  \item[] $D$ is the diameter of the Litz wire in meters;
  \item[] $\Delta_s$ is the distance between adjacent turns of the solenoid in meters;
  \item[] $\alpha$ is the twist angle of the helix formed by the strands inside the Litz wire;
  \item[] $G(z)$, $F(z)$ and $u_n$ are functions that describe the losses due to skin and proximity effects;
  \item[] $z=\frac{d}{\sqrt{2}}\delta$, where $\delta$ is the skin depth and depends on frequency.
\end{itemize}
\begin{figure}
  \centering
  \begin{subfigure}[b]{0.4\textwidth}
    \centering
    \includegraphics[width=\textwidth]{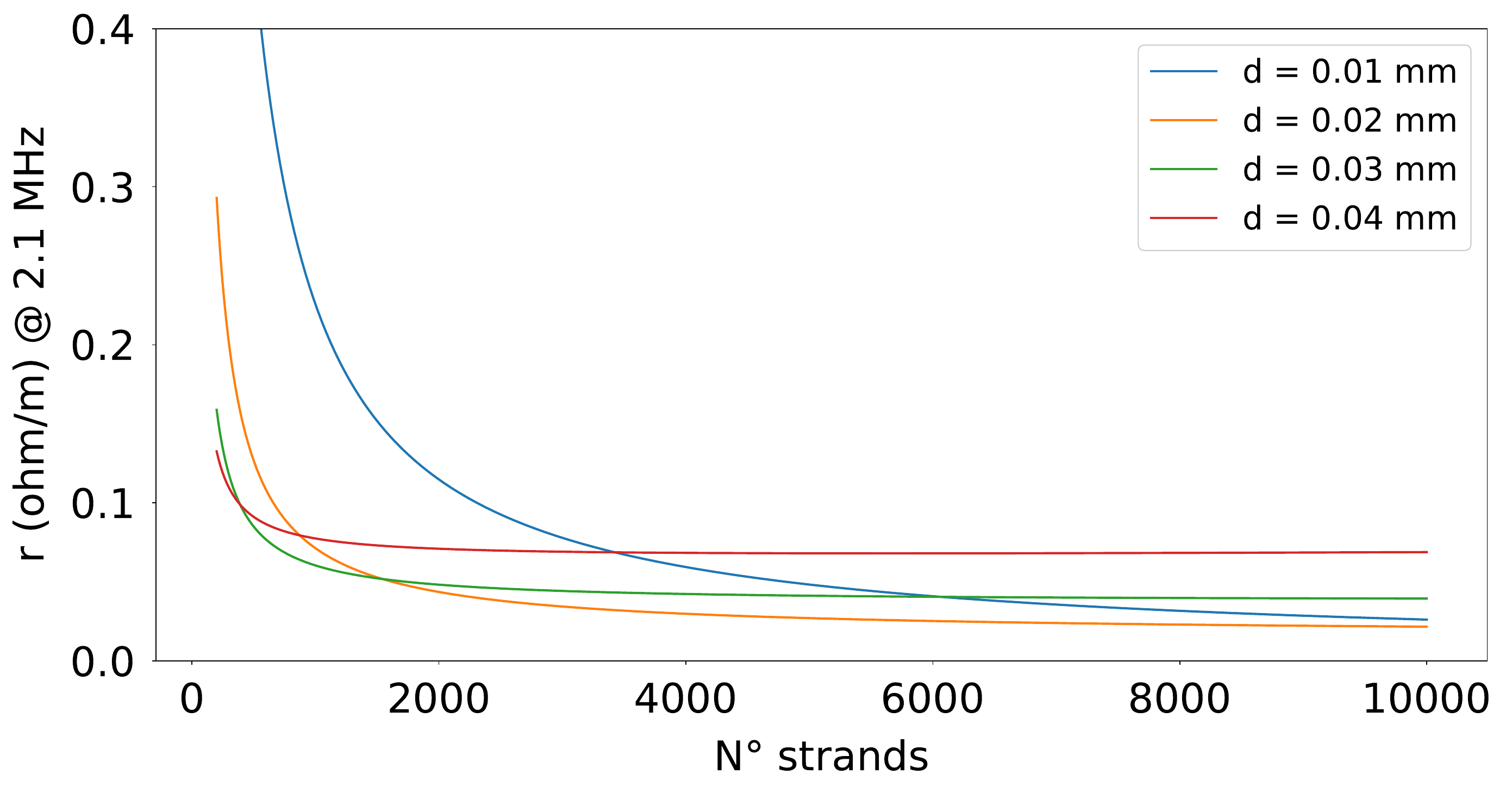}
    \caption{}
    \label{fig:litzSelection}
  \end{subfigure}
  \hspace{10pt}
  \begin{subfigure}[b]{0.4\textwidth}
      \centering
      \includegraphics[width=\textwidth]{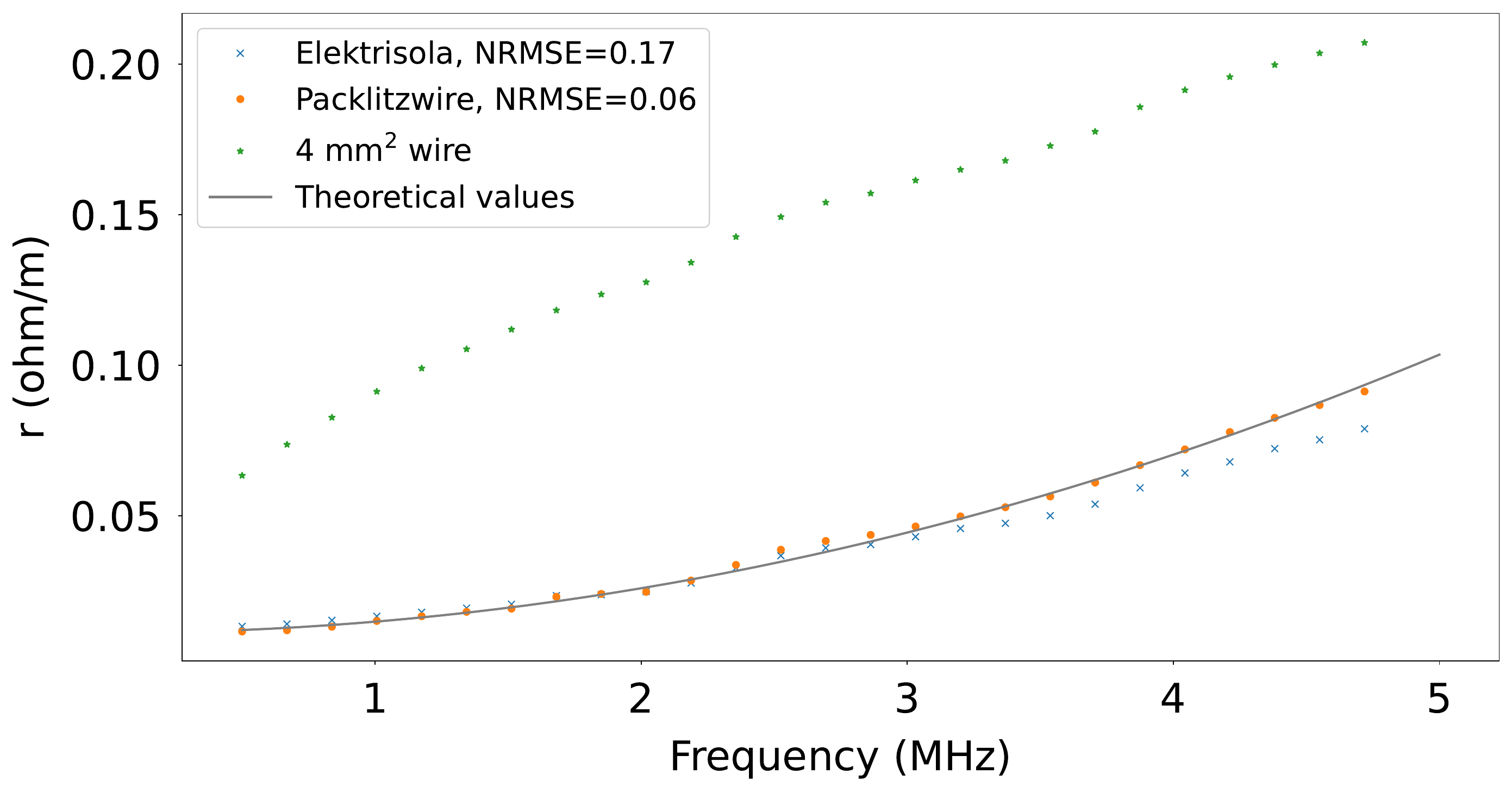}
      \caption{}
      \label{fig:litzMeasurements}
  \end{subfigure}
  \caption{(a): Theoretical values of the resistance per unit length provided at \SI{2.1}{\mega\hertz} for Litz wires with different strand diameters as a function of the number of strands. (b): Comparison between the theoretical and measured resistance per unit length. Measurements are performed on Litz wires provided by two different vendors with 5000 strands of \SI{0.02}{\milli\meter} diameter. The \gls{nrmse} are reported in the figure legend. For reference, the same measurement is shown for a solid \SI{4}{\milli\meter\squared} wire. All the measurements are acquired with an Agilent 4294A impedance analyzer }
\end{figure}

\figurename\,\ref{fig:litzSelection} implemented \ref{eq:litzTheoretical} to show the effect of the number and diameter of strands on the resistance per unit length for a \num{14} turns solenoid with \SI{125}{\milli\meter} length. After having verified that the results did not significantly change decreasing the number of \gls{rf} coil turns down to \num{8}, copper Litz wire made of \num{5000} strands with \SI{0.02}{\milli\meter} diameter was selected. \figurename\,\ref{fig:litzMeasurements} compares the \gls{ac} resistance per unit length computed through \eqref{eq:litzTheoretical} with values measured on two Litz wires provided by two different vendors: Elektrisola\cite{bib:elektrisola} and PACK LitzWire\cite{bib:packlitzwire}. Both wires were made of \num{5000} strands with \SI{0.02}{\milli\meter} diameter. The close agreement between the curves supports the reliability of the analytical model \eqref{eq:litzTheoretical} and the four times higher resistance measured for a solid wire with a comparable \SI{4}{\milli\meter\squared} section clearly justifies the adoption of the Litz wire to decrease the \gls{rf} coil losses. 

The \gls{em} simulations used for the design of the \gls{rf} coil modeled the windings as lossless \gls{pec}. Losses were accounted for in a second stage as a series resistance whose value was computed with \eqref{eq:litzTheoretical} and was added to the 1-port \gls{rf} coil model with CoSimPy\cite{bib:zanovello1}. Simulations included a simplified model of a human head positioned inside the \gls{rf} coil\cite{bib:3Dhead}. The head was simulated as a homogeneous medium with an electrical conductivity of \SI{0.5}{\siemens\per\meter} and relative electric permittivity of \num{80}. Even though these values did not match realistic properties of human tissues at \SI{2}{\mega\hertz}\cite{bib:itisDB} they corresponded to the measured properties of a tissue simulating liquid available in the authors' laboratory and used for the first benchtop measurements with the Vector Network Analyzer. All the quantities in \eqref{eq:optObjective} were computed inside the head model within a \SI{10}{\centi\meter} slab starting from the top of the head.

Once the optimization process identified the best coil configuration, a final two-port simulation was performed for computing the tuning and matching capacitor values. One port was connected to the coil supply and a second segmented the solenoid in two equal parts to decrease coil losses and the influence of loading on the resonance frequency of the circuit\cite{bib:decorps1}. CoSimPy was used to replace the first port with the matching network and the second with the tuning capacitor.

On the basis of the optimized coil design the housing was designed with the 3D CAD software FreeCAD\cite{bib:freecad} and printed with an Ultimaker S7 3D printer, using \gls{pla} and water-soluble \gls{pva} filaments for the structure and support, respectively. All \glspl{pcb} were designed with KiCad\cite{bib:kicad}.

\subsection{Benchtop Measurements}
\label{sec:benchtopMeasurements-Method}
The Q factor of the \gls{rf} coils were measured with two shielded loop probes\cite{bib:shieldedLoopProbe} connected to a \gls{vna} (INRIM: Copper Mountain \gls{vna} (mod. S5065), TU Graz/PTB: ZVL, Rohde \& Schwarz). The same \glspl{vna} were used to measure the S11 \SI{-3}{\decibel} bandwidth, \textit{i.e.}, the frequency bandwidth over which half of the incident power is reflected due to coil mismatch. Measurements were performed independently on the three \gls{rf} coil setup. Measurements at INRiM involved a \textit{SAM head} phantom\cite{bib:samPhantom} and the \textit{Hello World} phantom\cite{bib:osii_helloworld}, both filled with a characterized tissue simulating liquid (\SI{0.5}{\siemens\per\meter} electrical conductivity and \num{80} relative electric permittivity). TU Graz filled the \textit{Hello World} phantom with 1.5 g/l CuSO\textsubscript{4} distilled water. Since neither the \textit{SAM head} nor the \textit{Hello World} phantoms were available at PTB at the time of writing, measurements at PTB only involved the unloaded \gls{rf} coil.\\
The insertion loss of the \gls{rf} connector was measured at the connector reference planes through a Copper Mountain \gls{vna} (mod. S5065).

\subsection{EMI Coupling Analysis}
\label{sec:emiCoupling-Method}
\begin{figure}
    \centering
    \includegraphics[width=0.8\linewidth]{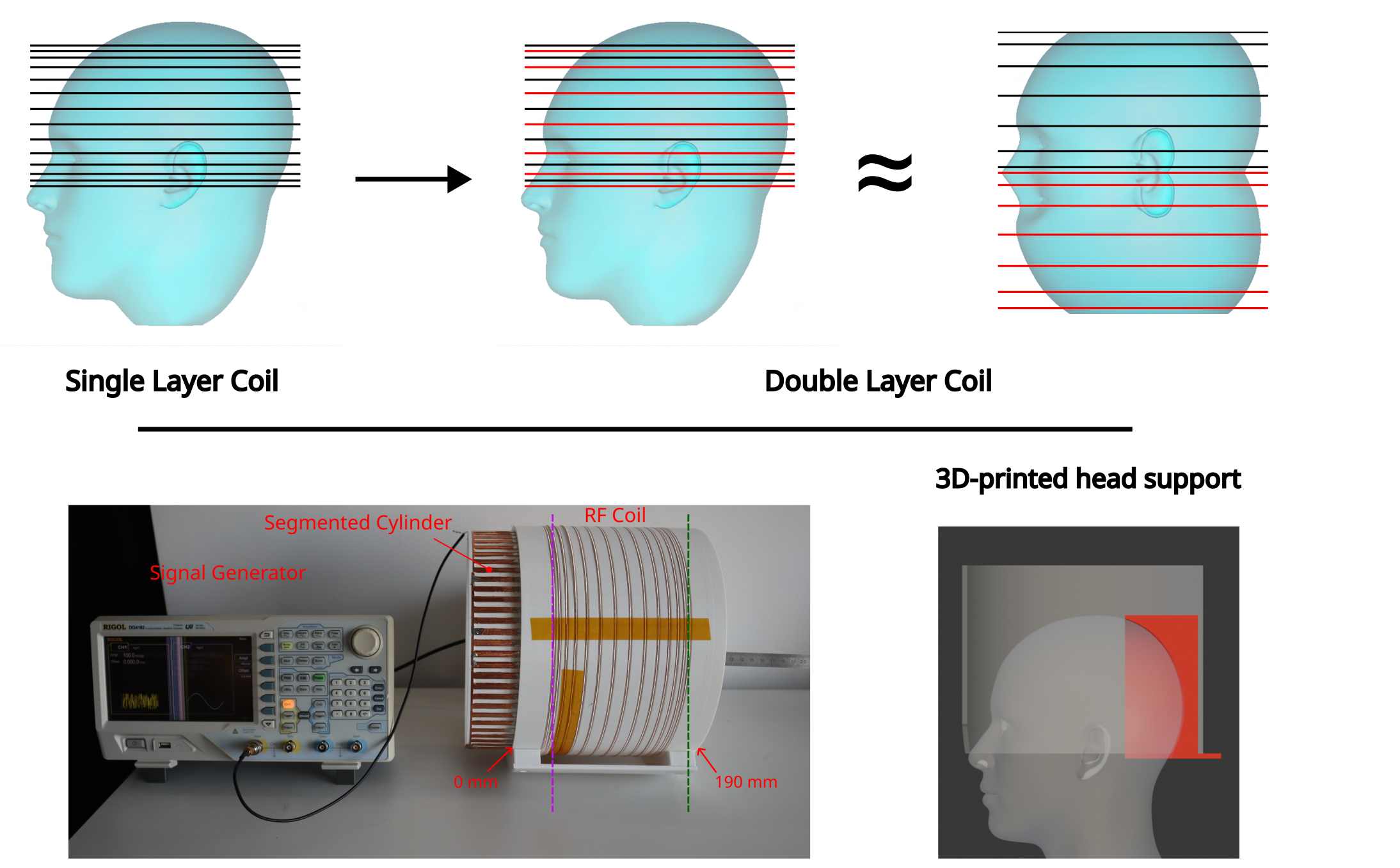}
    \caption{Top: Comparison between the conductor distribution for two winding configurations. For the double layer configuration, the solenoid turns are shown in black and red according whether they are winded from top to bottom or vice versa. The double layer configuration is expected to improve the longitudinal symmetry of the coil loading as sketched on the right side by unwrapping the two layer in a single equivalent one. Bottom: Setup used for the \gls{emi} analysis with the segmented copper cylinder photographed outside the magnet bore (left) and 3D rendering of the 3D printed support used for positioning the volunteer's head during in-vivo noise measurements (right).}
    \label{fig:doubleLayerSetup}
\end{figure}
Noise coupling as been shown to be a major issue for in-vivo imaging\cite{bib:Pfitzer2026, bib:lena1}. The human body, acting as an antenna\cite{bib:li1}, couples the environmental noise\cite{bib:guallartnaval1} with the \gls{rf} coil via capacitive coupling. He et al.\cite{bib:he1} showed how any load-coil asymmetry contributes to transform the common-mode voltage coupled through the human body into differential-mode voltage to the receiver. In the case of a head \gls{rf} coil, such asymmetry happens mainly along the main axis of the solenoid. The paper compares the noise performance of two \gls{rf} coil configurations. The configurations changed with respect to how the wire was winded along the solenoid main axis as \figurename\,\ref{fig:doubleLayerSetup} shows. The first configuration, named single layer coil, is the configuration referred in all the other section of the paper and corresponds to the wire winded from top to bottom and back to top with a straight wire parallel to the main coil axis. In the double layer configuration, the wire was winded first from top to bottom and then in the opposite direction. As \figurename\,\ref{fig:doubleLayerSetup} shows, by conceptually unwinding the double layer coil in a single layer equivalent one, this is expected to symmetrize the coupling of the coil with respect to the head and consequently reducing the noise coupling. In both configurations the number of turns was kept equal to \num{14}.

The noise coupling were assessed in two ways. First a segmented copper cylinder was used as load and connected to a signal generator (DG4162, RIGOL Technologies, China) in noise mode (see \figurename\,\ref{fig:doubleLayerSetup} bottom left). This cylinder was used in order to not significantly detune the coil. The cylinder was then inserted into the \gls{rf} coil from the top until it was completely inside. The noise coupled through the \gls{rf} coil was measured with the MaRGE\cite{bib:Algar_n_2024} relevant \textit{Noise} feature each \SI{10}{\milli\meter} step. The measurement was performed inside the magnet with the \gls{rf} coil positioned as it would be in a typical \gls{mri} acquisition. This measurements mimicked the analysis of Figure 7 in the He et al. paper\cite{bib:he1}, however with a mostly symmetrical structure. 
A second measurement was performed by measuring the noise coupling when a volunteer's head was positioned inside the coil. A 50 kHz wide noise spectrum was acquired for both coil configurations by averaging 1000 noise scans. Additionally, a noise factor was calculated as the ratio of the \gls{rms} of a single noise measurement and a \SI{50}{\ohm} baseline.
The 3D-printed support shown in \figurename\,\ref{fig:doubleLayerSetup} guaranteed repeatable results and avoided introducing an axial asymmetry due to accidental tilting of the head during the noise acquisition.

\subsection{Safety assessment}
\label{sec:safetyAssesment-Method}
The \gls{sar} was determined based on \gls{em} simulations in Sim4Life\cite{bib:s4l} with the human voxel model\cite{iacono1}. In addition, simulations with the same \gls{rf} coil setup were performed with an empty \gls{rf} coil, but inside the \gls{rf} shield. To validate the simulations, calibrated \Bop \gls{rms} measurements were performed. For this purpose, an open-source H-field pick-up probe\cite{bib:shieldedLoopProbe} was positioned within a TEM cell\cite{bib:Klepsch1}, which was connected to a function generator (SDG1032X Plus, Siglent) and an \gls{rf} power amplifier (Barthel HF-Technik). The output signal of the TEM cell and the pick-up coil were then measured with an oscilloscope, which allowed to calibrate the pick-up coil measurements with the generated H-field by the TEM cell. After calibration, the \Bop \gls{rms} was measured at the center of the empty \gls{rf} headcoil within the \gls{rf} shield using the calibrated H-field probe. 

\subsection{RF Connector}
\label{sec:rfConnector-Method}
An \gls{rf} connector was designed to connect the head coil to the OSI\textsuperscript{2} ONE scanner and to provide an open-source \gls{rf} connector platform that can be used to connect different \gls{rf} coils. The following list of requirements guided the design of the connector:
\begin{itemize}
  \item 1x coaxial contact capable of carrying at least 1 kW of peak \gls{rf} power;
  \item 36x contacts to provide \gls{dc} and logic signals to on-coil hardware such as PIN diodes, Q-damping circuit, thermal sensors, etc.;
  \item Minimal distortion of the static magnetic field;
  \item Sturdy 3D printable housing with the following characteristics:
  \begin{itemize}
    \item Keyed connector design to prevent incorrect mating;
    \item Smooth, easy to clean and non-conductive surfaces;
    \item Reliable alignment between all the contacts;
    \item Locking mechanism to maintain contact in case of vibrations or minor accidental force on the cables;
    \item Easy and quick detaching mechanism with one hand in case of emergency;
    \item Dust cover to protect exposed contacts when the connector is detached;
  \end{itemize}
  \item Uses off-the-shelf parts available from multiple manufacturers;
  \item Embedded I\textsuperscript{2}C-EEPROM to store \gls{rf} coil information in compliance with the \gls{mrcods}\cite{bib:mrcods}
\end{itemize}
The connector housing and \glspl{pcb} were designed using FreeCAD\cite{bib:freecad} and KiCad\cite{bib:kicad}, respectively. The housing was printed with an Ultimaker S7 3D printer, using \gls{pla} and water-soluble \gls{pva} filaments for the structure and support, respectively.

\subsection{MRI Experiments}
\label{sec:mriExperiments-Method}

All \gls{mri} experiments were performed with the TU Graz \gls{mri} scanner. The scanner was built locally and uses the open-source OSI\textsuperscript{2} ONE design \cite{bib:osii_one}. It utilizes a MaRCOS-based console \cite{bib:Negnevitsky_2023} \cite{bib:osii_console} with MaRGE\cite{bib:Algar_n_2024} as console software, which provides various pulse sequences and hardware calibration options.

Both gradient and \gls{rf} amplifiers are based on open-source projects \cite{bib:osii_rfpa} \cite{bib:osii_gpa}. The system uses a permanent magnet array \cite{bib:osiiMagnet} with a field strength of about \SI{49}{\milli\tesla} (\SI{2.11}{\mega\hertz} at \SI{22}{\celsius}), shimmed to a homogeneity of \SI{3100}{ppm} over a \SI{200}{\milli\meter} \gls{dsv}\footnote{\Bz homogeneity was evaluated as $(B_{0,\rm max}-B_{0,\rm min})/B_{0,\rm avg}$}. The gradient coils (x: \SI{0.438}{\milli\tesla\per\meter\per\ampere}, y: \SI{0.910}{\milli\tesla\per\meter\per\ampere}, z: \SI{0.597}{\milli\tesla\per\meter\per\ampere}) were built in-house.

The active \gls{txrx} switch was built following an open-source design from OCRA \cite{bib:a4im_txrx}, while while a self-built \gls{lna} was used (Gain=\SI{44}{\decibel} at \SI{2.11}{\mega\hertz}).

Due to \gls{emi} coupling via the body, an inner \gls{rf} shield\cite{bib:Pfitzer2026}, named FENCE,  was added for \gls{emi} mitigation during in-vivo measurements. This resulted in a reduction of the Q factor of the \gls{rf} coil to \num{292} in the loaded and \num{350} in the unloaded configuration.

Since the designed \gls{rf} connector was not available at TU Graz at the time of experiments, the \gls{rf} coil connected to the scanner through a standard SMA connector.

Prior to all \gls{mri} experiments, the flip angle and Larmor frequency were calibrated using MaRGE's calibration sequences. Noise levels were compared to a baseline configuration where the \gls{txrx} switch was terminated with \SI{50}{\ohm}. These noise measurements were repeated under loaded conditions for both phantom and in-vivo experiments to ensure no significant additional \gls{emi} was introduced.

First-order shimming was performed using the gradient coils. Due to eddy currents on the \gls{rf} shield positioned between the gradient coils and \gls{rf} coil, which caused a shift in k-space center, the readout window and gradient timing were adjusted to center the k-space before measurements.

A 3D Rapid Acquisition with Relaxation Enhancement (RARE) sequence. Tables \ref{tab:mri_params_phantom} and \ref{tab:mri_params_invivo} collect the sequence parameters used for the phantom and in-vivo scanning, respectively. 

\begin{table}[!ht]
\centering
\caption{Phantom \gls{mri} Acquisition Parameters}
\label{tab:mri_params_phantom}
\begin{tabular}{ll}
\hline
\textbf{Parameter} & \textbf{Value} \\
\hline
Sequence & 3D RARE \\
Flip Angle & 90$^\circ$ 180$^\circ$\\
TR/echo spacing/echo train length & \SI{500}{\milli\second} / \SI{20}{\milli\second} / \num{5} \\
Field-of-view & \qtyproduct{120x120x120}{\milli\meter}\\
Data points & \numproduct{60x60x60} (RO, PH1, PH2)\\
Voxel size & \qtyproduct{2x2x2}{\milli\meter}\\
Acquisition bandwidth & \SI{10}{\kilo\hertz}\\
\gls{rf} pulse length & \SI{210}{\micro\second} / \SI{420}{\micro\second}\\
Trajectory & Cartesian Inside-out \\
Averages & \num{1}\\
Acquisition Time & \SI{6}{\minute}\\
\hline
\end{tabular}
\end{table}

\begin{table}[ht!]
\centering
\caption{In-Vivo \gls{mri} Acquisition Parameters}
\label{tab:mri_params_invivo}
\begin{tabular}{ll}
\hline
\textbf{Parameter} & \textbf{Value} \\
\hline
Sequence & 3D RARE \\
Flip Angle & 90$^\circ$ 180$^\circ$ \\
TR/echo spacing/echo train length & \SI{500}{\milli\second} / \SI{20}{\milli\second} / \num{5}\\
Field-of-view & \qtyproduct{240x200x150}{\milli\meter} \\
Data points & \numproduct{120x100x30} (RO, PH1, PH2)\\
Voxel size & \qtyproduct{2x2x5}{\milli\meter}\\
Acquisition bandwidth & \SI{15}{\kilo\hertz}\\
\gls{rf} pulse length & \SI{250}{\micro\second} / \SI{500}{\micro\second}\\
Trajectory & Cartesian Inside-out \\
Averages & \num{1} \\
Acquisition Time & \SI{5}{\minute}\\
\hline
\end{tabular}
\end{table}

Images were reconstructed using an inverse Fast Fourier Transform without any additional filtering or corrections for \Bz inhomogeneities or gradient imperfections. 

In-vivo experiments were carried out with approval from the TU Graz ethics committee and written informed consent was obtained from the subject prior to the measurements.

\section{Results}
\subsection{\gls{rf} Coil Design}
\label{sec:rfCoilDesign-Results}
\begin{figure}
  \centering
  \includegraphics[width=.9\textwidth]{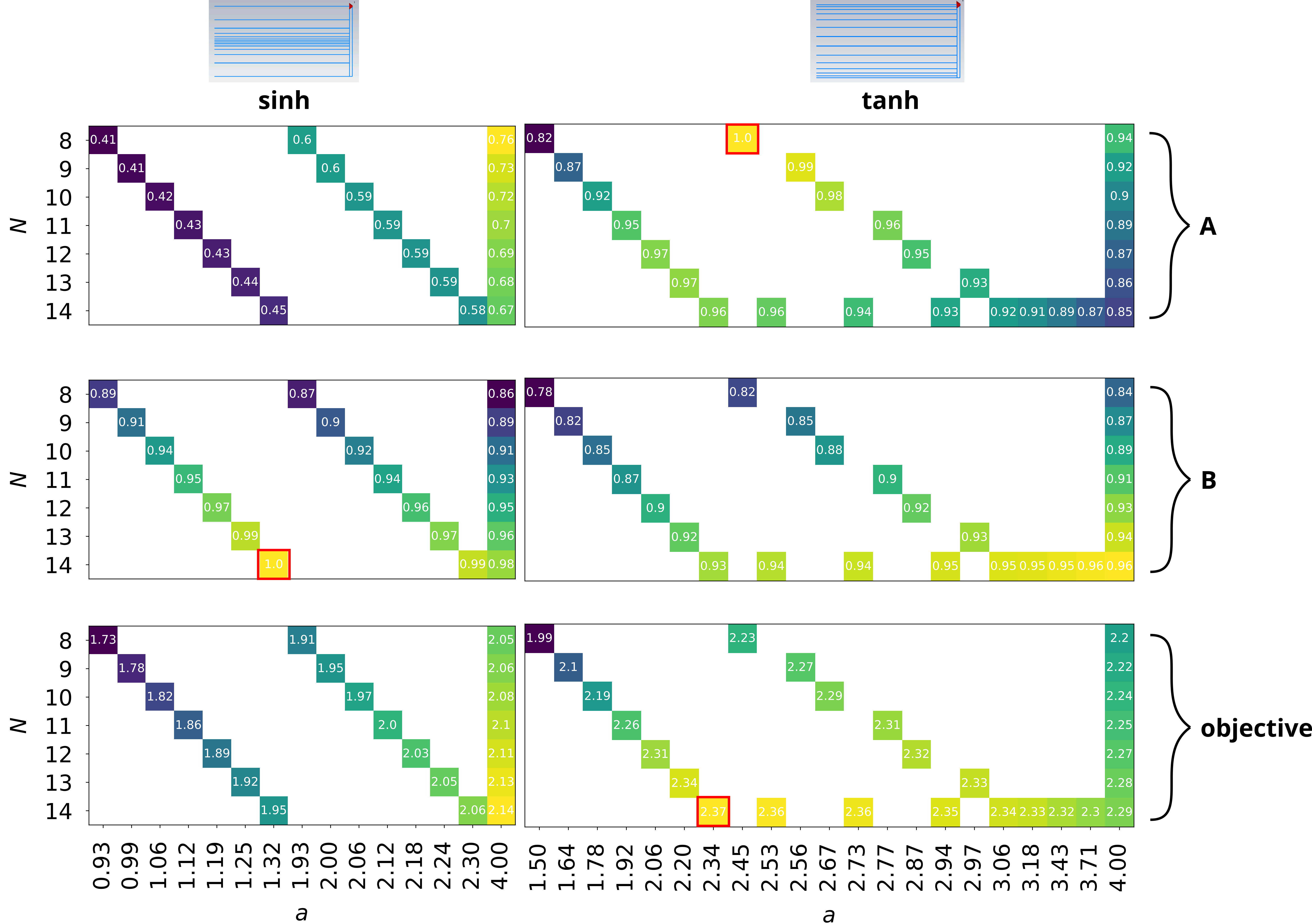}
  \caption{Values of A, B and objective with reference to \eqref{eq:optObjective} for the \textit{sinh} and \textit{tanh} conductors distribution. The values of the parameters are reported as a function of the number of turns of the solenoid ($N$) and the $a$ factor. The red squares identify the best configuration for a specific parameter.}
  \label{fig:optResults}
\end{figure}
\figurename\,\ref{fig:optResults} collects the results of the \num{48} simulations. Whereas the \textit{tanh} conductors distribution leads to the best \Bop homogeneity, the \textit{sinh} distribution results in the best sensitivity. Despite the \num{1.5} multiplying factor of $B$ in \eqref{eq:optObjective}, the \textit{tanh} distribution provides the best value of the objective, for $a=2.34$ and $N=14$. Interestingly and somewhat counterintuitively, the \Bop homogeneity ($A$) slightly worsens as the number of turns increases. This result seems to be consistent with simulations and measurements made by other authors on a regular solenoid\cite{bib:blasiak1}. As regards the \gls{rf} coil sensitivity ($B$), it increases with the number of turns. This was expected since, as long as proximity effects are limited, both the \Bop and power increase with N. 

\begin{figure}
  \centering
  \begin{subfigure}[b]{0.4\textwidth}
    \centering
    \includegraphics[width=\textwidth]{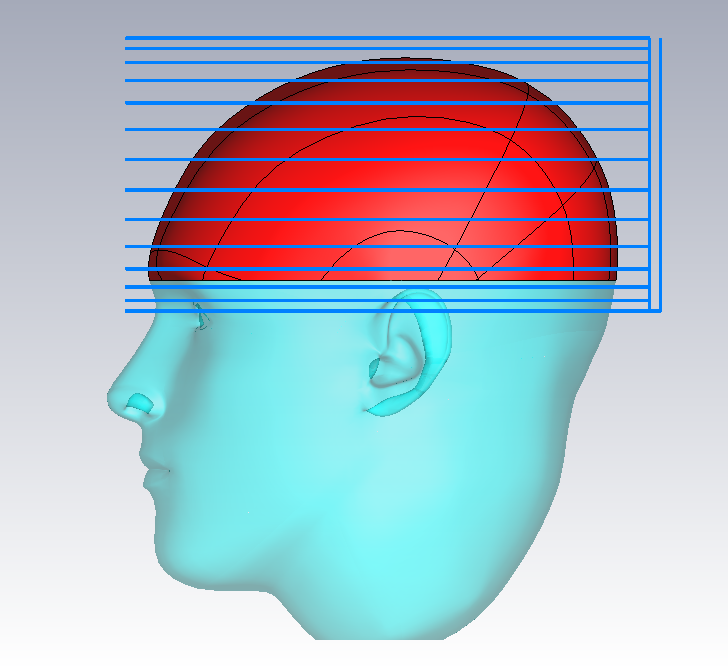}
    \caption{}
    \label{fig:actualConductorDistrib}
  \end{subfigure}
  \hspace{10pt}
  \begin{subfigure}[b]{0.4\textwidth}
    \centering
    \begin{tabular}{ l c c c c}
      \multirow{2}{8em}{Shield Diameter} & \multicolumn{4}{c}{Capacitance}\\ 
      & $C_1$ & $C_t$ & $C_2$ & $C_3$\\
      \hline\\
      \SI{280}{\milli\meter} & \SI{322}{\pico\farad} & \SIrange{1}{23}{\pico\farad} & \SI{1}{\nano\farad} & \SI{300}{\pico\farad}\\
      \SI{284}{\milli\meter} & \SI{225}{\pico\farad} & \SIrange{2}{120}{\pico\farad} & \SI{1}{\nano\farad} & \SI{268}{\pico\farad}\\
      \SI{290}{\milli\meter} & \SI{280}{\pico\farad} & \SIrange{1}{23}{\pico\farad} & \SI{1}{\nano\farad} & \SI{280}{\pico\farad}\\
    \end{tabular}
    \vspace{5pt}

    \includegraphics[width=\textwidth]{equivalentCircuit.tikz}
    \caption{}
    \label{fig:coilEquivCircuit}
  \end{subfigure}\\
  \vspace{10pt}
  \begin{subfigure}[b]{0.4\textwidth}
      \centering
      \includegraphics[width=\textwidth]{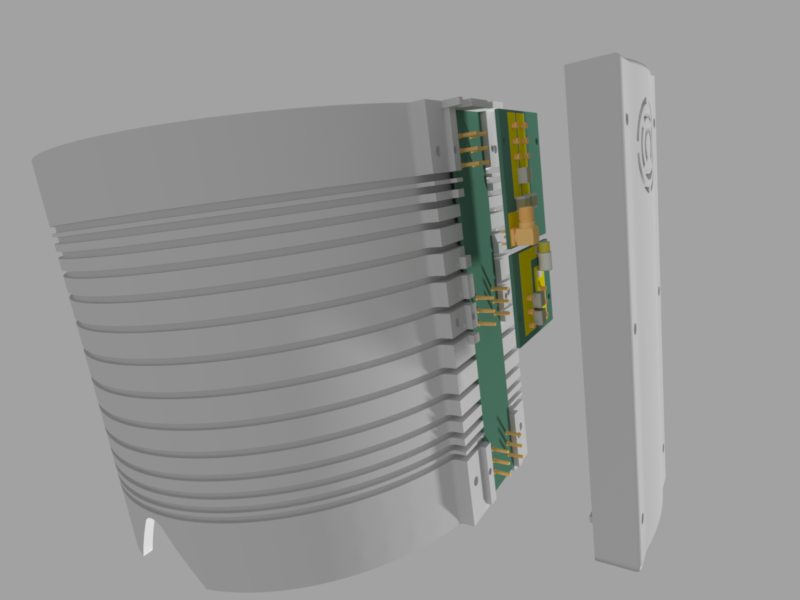}
      \caption{}
      \label{fig:coilRender}
  \end{subfigure}
  \hspace{10pt}
  \begin{subfigure}[b][][c]{0.4\textwidth}
      \centering
      \includegraphics[width=\textwidth]{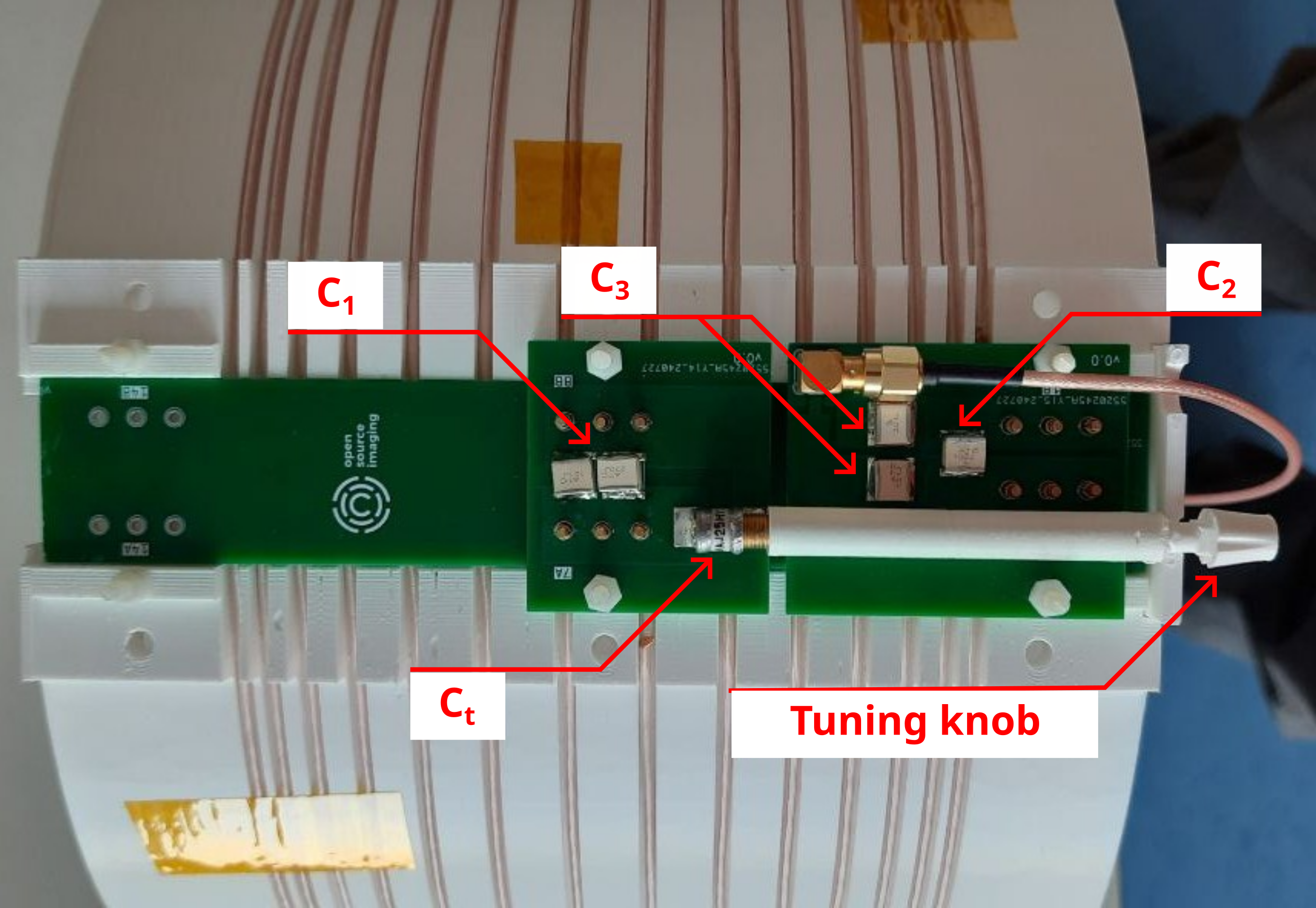}
      \caption{}
      \label{fig:coilPhoto}
  \end{subfigure}
  \caption{(a): Optimized conductor distribution ($a=2.34, N=14$) with the head model centered inside the coil. The averaging volume used in \eqref{eq:optObjective} is the red volume of the head. (b): Equivalent circuit of the coil showing the tuning and matching capacitors optimized for the three \gls{rf} coil setup. $R$ and $L$ model the resistance and inductance of the coil, respectively. (c): 3D rendering of the coil housing showing the detachable \glspl{pcb} hosting the tuning and matching circuitry. (d): Photo of the \gls{rf} coil with the exposed \glspl{pcb} and tuning knob.}
\end{figure}

\figurename\,\ref{fig:actualConductorDistrib} shows the optimum conductor distribution. The head model is positioned inside the \gls{rf} coil showing in red the averaging volume used in \eqref{eq:optObjective}. \figurename\,\ref{fig:coilEquivCircuit} shows the equivalent circuit of the \gls{rf} coil together with the tuning and matching capacitors optimized for the three \gls{rf} coil setup. The capacitor $C_1$ segments the \gls{rf} coil in two equal parts and a variable capacitor $C_t$ is added in parallel to $C_1$ to fine-tune the coil. A symmetrical circuit, made of one parallel ($C_2$) and two equal series ($C_3$) capacitors, matches the \gls{rf} coil to \SI{50}{\ohm} at the \SI{2.04}{\mega\hertz} Larmor frequency of the built magnet. The \Bz value measured on the built magnet array was \SI{48}{\milli\tesla} and differed from the \SI{50}{\milli\tesla} reference value\cite{bib:osiiMagnet}. This justifies the difference between the actual \SI{2.04}{\mega\hertz} and design \SI{2.1}{\mega\hertz} \gls{rf} coil frequencies. Whereas such a difference impacts on the value of the tuning and matching capacitors, it is not expected to affect the optimization results. All fixed capacitors used in the INRiM \gls{rf} coil are of high Q, AVX 100C series type and the variable capacitor is a Voltronics A\_25 HV. The table in \figurename\,\ref{fig:coilEquivCircuit} collects the capacitance of the capacitors. Including all capacitor losses in the \gls{rf} coil model, simulations predicted a coil sensitivity $\psi$ of \SI{17.4}{\micro\tesla\per\sqrt\watt} and a normalized standard deviation $\frac{\sigma(B_1)}{\overline{B_1}}$ of \num{0.12}.

\figurename\,\ref{fig:coilRender} and \figurename\,\ref{fig:coilRender} show a rendering and a photo of the designed housing. The housing presents ruts for the correct positioning of the conductor. An insert for the nose helps with the head positioning inside the coil and decreases the claustrophobia effect by facilitating breathing. The conductors are soldered on the back of a PCB on which other smaller \glspl{pcb} are connected by means of multiple pins. This solution allows for an easier maintenance, update and testing of different components on the same \gls{rf} coil, \textit{e.g.}, addition of PIN diodes, implementation of Q spoiling circuit, etc. A 3D printed knob and shaft hosts a screwdriver head for an easy tuning of the coil by trimming the variable capacitor. The overall dimensions of the \gls{rf} coil housing are \qtyproduct{243 x 243 x 190}{\milli\metre}.

\subsection{Benchtop Measurements}
\label{sec:benchtopMeasurements-Results}
\begin{table}
  \centering
  \caption{Results of benchtop measurements performed independently on three assemblies of the \gls{rf} coil.}
  \label{tab:rfCoilBechtop}
  \begin{tabular}{l c c c c c c c}
    \multirow{2}{4em}{\gls{rf} Coil} & \multirow{2}{4em}{Phantom} & \multicolumn{2}{c}{Shield} & \multicolumn{2}{c}{$Q$} & \multicolumn{2}{c}{S11 \SI{-3}{\decibel} bandwidth}\\[3pt]
    & & diameter & thickness & unloaded & loaded & unloaded & loaded\\[3pt]
    \hline\\[3pt]
    \multirow{2}{4em}{INRiM} & SAM head & \SI{280}{\milli\meter} & \SI{0.3}{\milli\meter} & 500 & 360 & \SI{9.7}{\kilo\hertz} & \SI{11.5}{\kilo\hertz}\\
    & Hello World & \SI{290}{\milli\meter} & \SI{0.3}{\milli\meter} & 570 & 530 & \SI{8.65}{\kilo\hertz} & \SI{9}{\kilo\hertz}\\[3pt]
    TU Graz & Hello World & \SI{284}{\milli\meter} & \SI{1.0}{\milli\meter} & 444 & 435 & \SI{10.5}{\kilo\hertz} & \SI{10.7}{\kilo\hertz}\\[3pt]
    PTB  & - & \SI{290}{\milli\meter} &  \SI{0.15}{\milli\meter} & 306 & - & \SI{10.5}{\kilo\hertz}& -\\
  \end{tabular}
\end{table}
\tablename\,\ref{tab:rfCoilBechtop} collects the results of the benchtop measurements performed independently on the three different assemblies of the designed \gls{rf} coil with different shield diameters and thicknesses. Focusing on INRiM measurements, as expected larger shield diameters led to higher Q factors and lower bandwidths. While the \textit{Hello World} phantom did not significantly decrease the unloaded Q factor of the coil, the \textit{SAM Head} had a higher impact due to its larger dimensions. Both TU Graz and PTB coils showed lower Q factors which cannot be explained solely by the different experimental setup. Possible explanations for this discrepancy include different capacitor Q factors, defective soldering or capacitors, or water trapped into the \gls{pla} \gls{rf} coil housing due to the water-bath used to remove the \gls{pva} 3D printing water-soluble support material. This last issue was found to be critical for the INRiM coil and required to heat the coil housing for about \qty{12}{\hour} at \qty{45}{\celsius} to make the water evaporate. A similar procedure could prove to be beneficial also for TU Graz and PTB coils.

The insertion loss of the \gls{rf} connector resulted to be better than \SI{0.1}{\decibel} up to \SI{300}{\mega\hertz} and lower than \SI{0.03}{\decibel} at \SI{2}{\mega\hertz}.

\subsection{EMI Coupling Analysis}
\label{sec:emiCoupling-Results}
\begin{figure}
  \centering
  \begin{subfigure}[b][][c]{0.8\textwidth}
    \centering
    \includegraphics[width=0.8\textwidth]{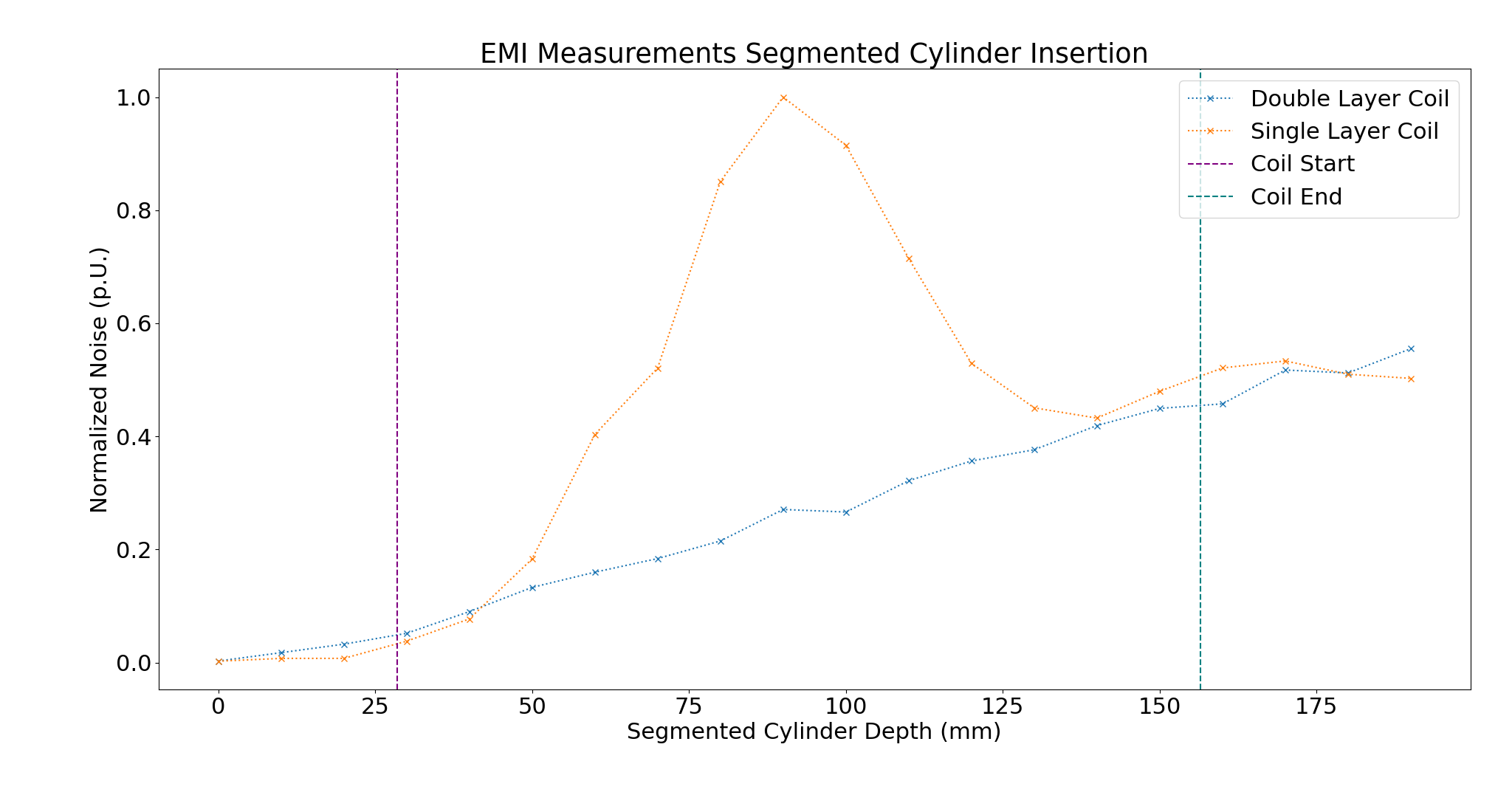}
    \caption{}
    \label{fig:fenceNoise}
\end{subfigure}\\
\begin{subfigure}[b][][c]{0.8\textwidth}
  \centering
  \includegraphics[width=0.8\textwidth]{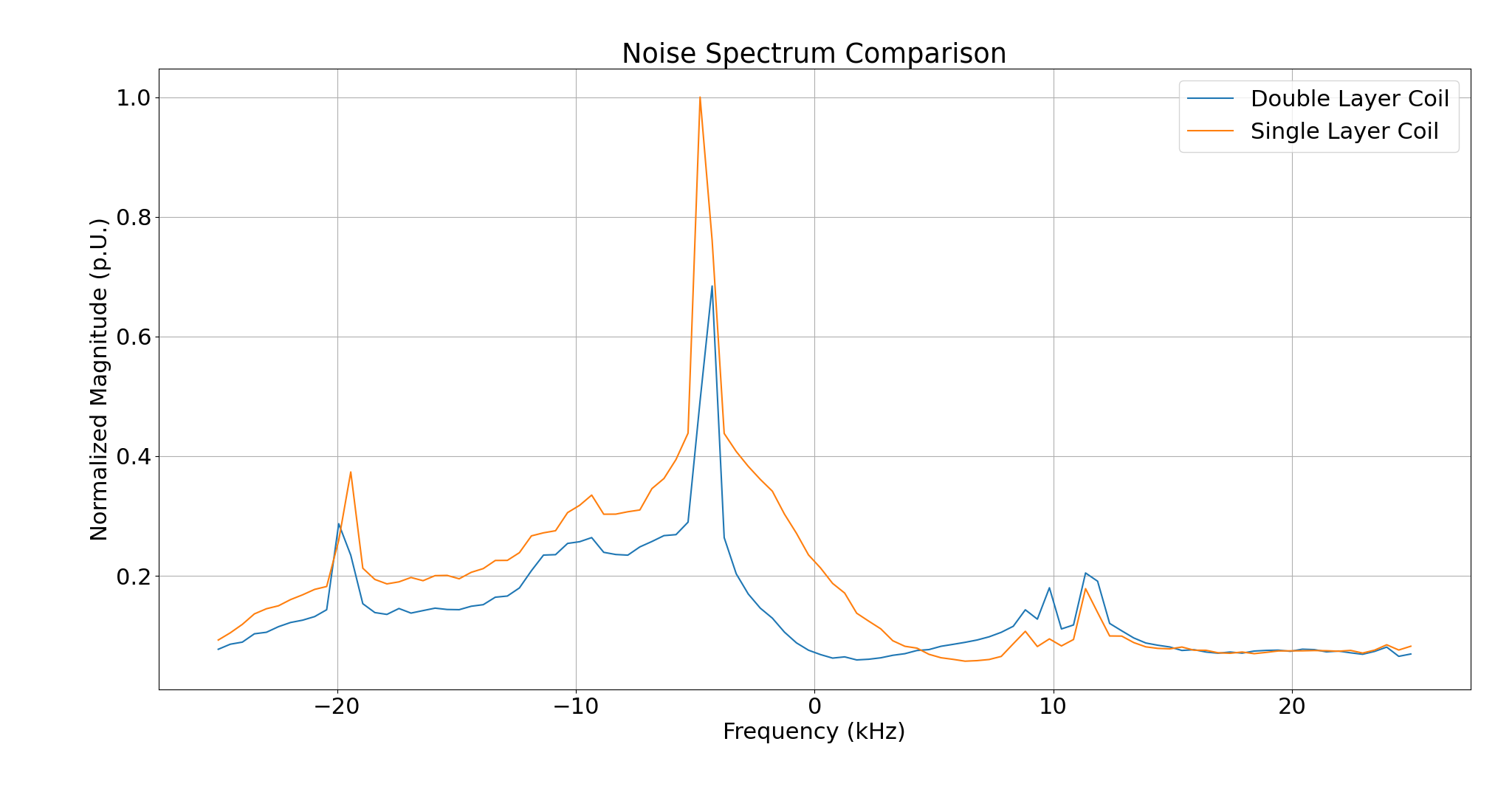}
  \caption{}
  \label{fig:headNoise}
\end{subfigure}
  \caption{(a): \gls{emi} Measurements with both coil configurations at different segmented copper cylinder insertion depths. The beginning and end of the \gls{rf} coil are marked with vertical lines (see also bottom left of \figurename\,\ref{fig:doubleLayerSetup}) (b): In-Vivo \gls{emi} coupling experiments.}
  \label{fig:noiseAnalysisResults}
\end{figure}

\figurename\,\ref{fig:noiseAnalysisResults} shows the results of the noise analysis and compares the behavior of the single and double layer coils. \figurename\ref{fig:fenceNoise} reports the noise acquired with the segmented copper cylinder used as load and connected to the noise signal generator. A significant increase in \gls{emi} coupling can be observed for the single layer coil for the cylinder being half inserted into the \gls{rf} coil. This was expected since this condition represents that with the maximum contemporary asymmetry and noise coupling and aligns well with the results of He at al.\cite{bib:he1}. The double layer instead exhibits a monotonic behavior of the noise amplitude as a function of the cylinder depth. While such an increase is justified by the increasing noise coupling between the cylinder and the \gls{rf} coil, interestingly the result does not show any relation between noise pick-up and coil asymmetry. This suggests that the double layer coil is effective in symmetrizing the load in the axial direction with a decrease in noise pick-up up to about \SI{75}{\percent}.
\figurename\ref{fig:headNoise} shows the results of the in-vivo \gls{emi} coupling. In this case the difference between the noise picked up by the two coils is less marked. The ratio of the \gls{rms} noise with respect to the \SI{50}{\ohm} baseline is \num{21.66} and \num{14.49} for the single and double layer coil, respectively. This suggest that for a less regular load, such as the human head, the conductor distribution along the solenoid axis is not fully effective in minimizing the noise coupling. In addition, despite the use of the head support, asymmetries along the minor coil axes may still have been present.

\subsection{Safety simulations}
\label{sec:safetyAssesment-Results}

\begin{figure}
    \centering
    \includegraphics[width=\linewidth]{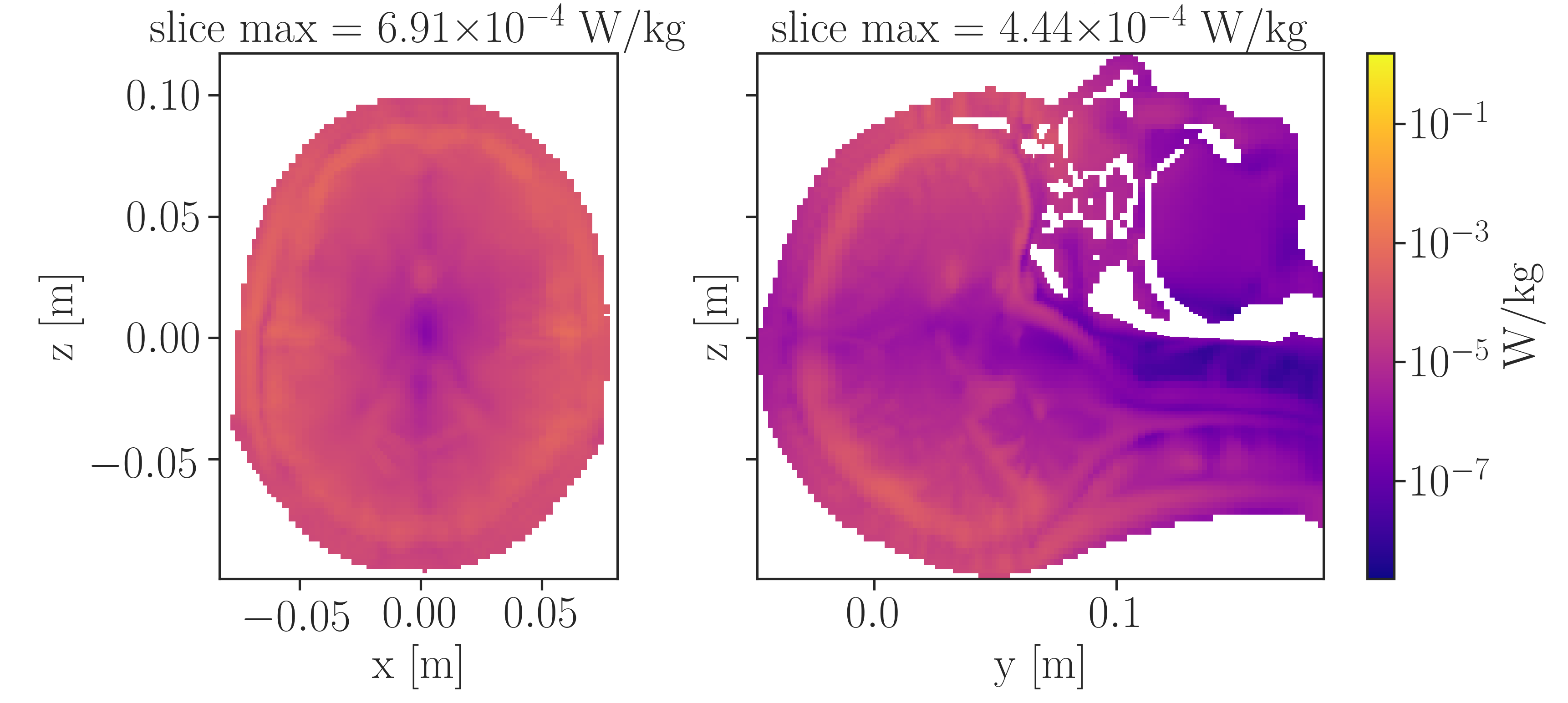}
    \caption{\SI{10}{\gram} averaged \gls{sar} profiles on representative axial and sagittal slices}
    \label{fig:10gSAR}
\end{figure}
The \gls{sar} simulation results are displayed in \figurename\,\ref{fig:10gSAR} in a representative axial and sagittal slice through the head. Overall the maximum \SI{10}{\gram} averaged local \gls{sar} in the human head is \SI{8.24e-4}{\watt\per\kilo\gram} at a \Bop \gls{rms} of \SI{1}{\micro\tesla}. The in-vivo sequence from \tablename\,\ref{tab:mri_params_invivo} uses one \SI{250}{\micro\second} and five \SI{500}{\micro\second} \gls{rf} pulses for \SI{90}{\degree} and \SI{180}{\degree} flip angles, respectively, each \SI{500}{\milli\second} TR. This results in an \gls{rms} \Bop of about \SI{1.74}{\micro\tesla} and a maximum \SI{10}{\gram} averaged local \gls{sar} of \SI{2.5}{\milli\watt\per\kilo\gram}. 

Despite this low \gls{sar} value, attention must be paid not to exploit all the power available from the \gls{rf} amplifier used for in-vivo imaging\cite{bib:osii_rfpa}, \textit{i.e.,} \SI{1}{\kilo\watt} with a \SI{10}{\percent} duty cycle. In this case, accounting for the simulated \gls{rf} coil sensitivity of \SI{17.5}{\micro\tesla\per\sqrt\watt}, the maximum \SI{10}{\gram} averaged local \gls{sar} scales to about \SI{25}{\watt\per\kilo\gram} exceeding the  IEC 60601:33 standard limits of \SI{20}{\watt\per\kilo\gram} in first level controlled operating mode\cite{bib:IEC60601-2-33-2022}.

\subsection{RF Connector}
\label{sec:rfConnector-Results}
\begin{figure}
  \centering
  \begin{subfigure}[b][][c]{0.6\textwidth}
    \centering
    \includegraphics[height=6cm]{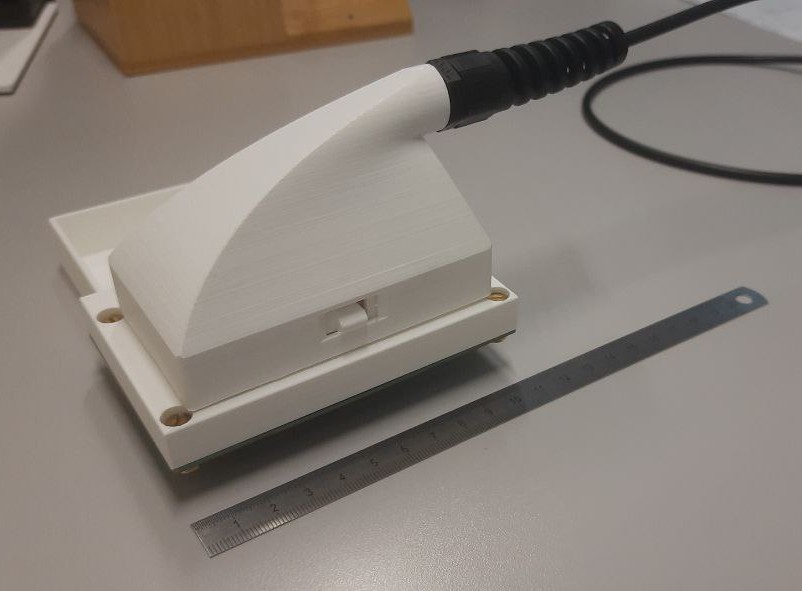}
    \caption{}
    \label{fig:connectorPhoto}
\end{subfigure}
\hspace{-1.5cm}
\begin{subfigure}[b][][c]{0.2\textwidth}
  \centering
  \includegraphics[height=6cm]{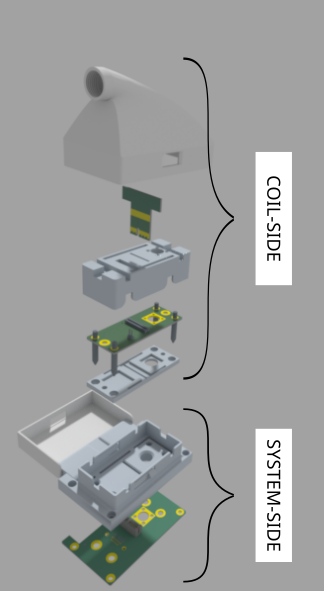}
  \caption{}
  \label{fig:explodedConnector}
\end{subfigure}
  \caption{(a): Photo of the mated \gls{rf} connector behind a \SI{20}{\centi\meter} ruler. (b) 3D rendering of the exploded \gls{rf} connector showing the multiple components.}
  \label{fig:rfConnector}
\end{figure}
\figurename\,\ref{fig:rfConnector} shows a photo of the \gls{rf} connector sided by a 3D rendering of the exploded model. The \gls{rf} contacts are a flanged version of the ubiquitous BNC connector where the bayonet-style nut is absent on the male side. The use of BNC connectors allow the coil to be directly connected to standard \gls{rf} test equipment and guarantees a continuous coaxial path which ensure excellent performance. Both connector halves are connected to the MR system and the \gls{rf} coil through SMA coaxial cables. \SI{1.6}{\milli\meter} thick FR4 \glspl{pcb} are used as frames on which contacts and other parts are mounted, and are fastened to the connector housing using non-magnetic screws. An edge card is mounted orthogonally to the connector plane and ensures \num{36} connections for \gls{dc} and logic signals. Its length is designed to ensure an optimum connection of all the contacts when the two halves of the connector are mated. In addition, the mating order of the traces on the edge card PCB can be adjusted by modifying their lengths. A Dsub9 connector provides the \gls{dc} and logic signals to an I\textsuperscript{2}C-EEPROM, leaving all the other \num{27} ($36-9$) connections either connected to the logic ground or unconnected. According to the \gls{mrcods} standard\cite{bib:mrcods}, one of the contacts, grounded on the coil-side half, is made available to the Dsub9 connector. This can be used to raise an interrupt on a pulled-up \glspl{gpio} each time the coil is connected or disconnected. 3D-printed plates protect the contacts and include ramps to assist the alignment of the two connector halves. Correct mating is further ensured by nonmagnetic guide pins. A cover, hinged to the system-side half, protects the circuitry when the coil-side half is not connected. The same mechanism used to lock the cover is also used to lock the coil-side half when connected.

\subsection{MRI Experiments}
\label{sec:mriExperiments-Results}
\begin{figure}[h!]
    \centering
    \includegraphics[width=1\linewidth]{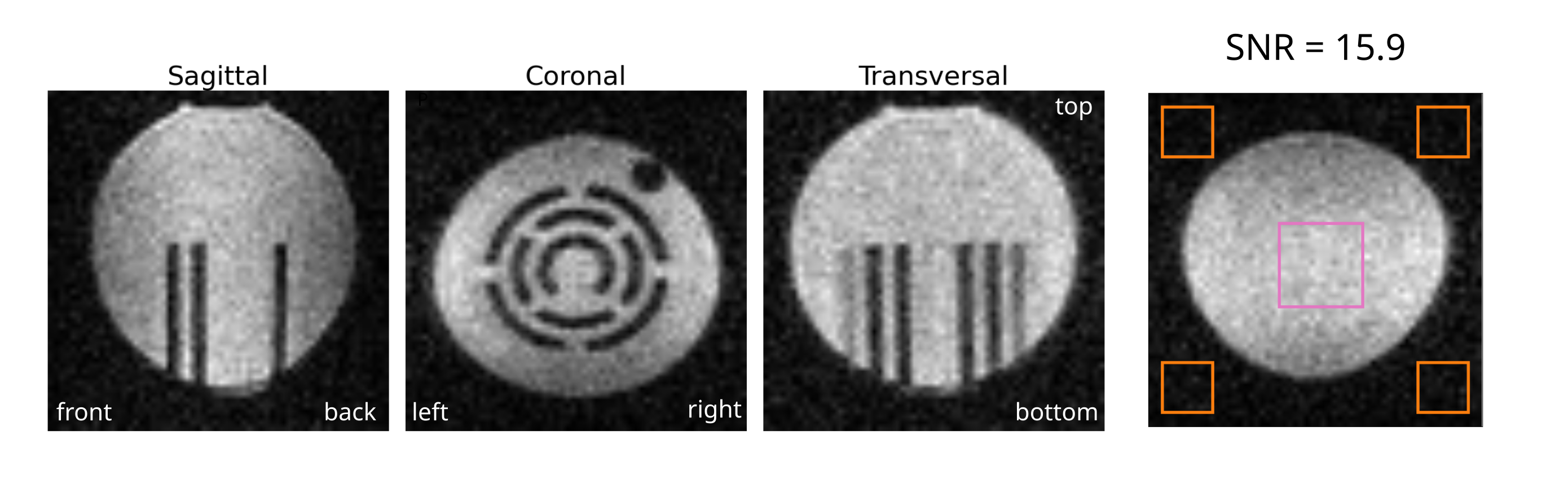}
    \caption{3D RARE images (\qtyproduct{2x2x2}{\milli\meter}) of the OSII \textit{Hello World} Phantom (left). 
Single slice (right) of the phantom. \gls{snr} was calculated using four noise ROIs placed outside the imaging region (orange) and a central ROI for signal mean estimation (pink). \gls{snr} was calculated as the ratio of the mean signal and the standard deviation of the noise ROIs.}
    \label{fig:phantom-results-tugraz}
\end{figure}

\begin{figure}[h!]
    \centering
    \includegraphics[width=\linewidth]{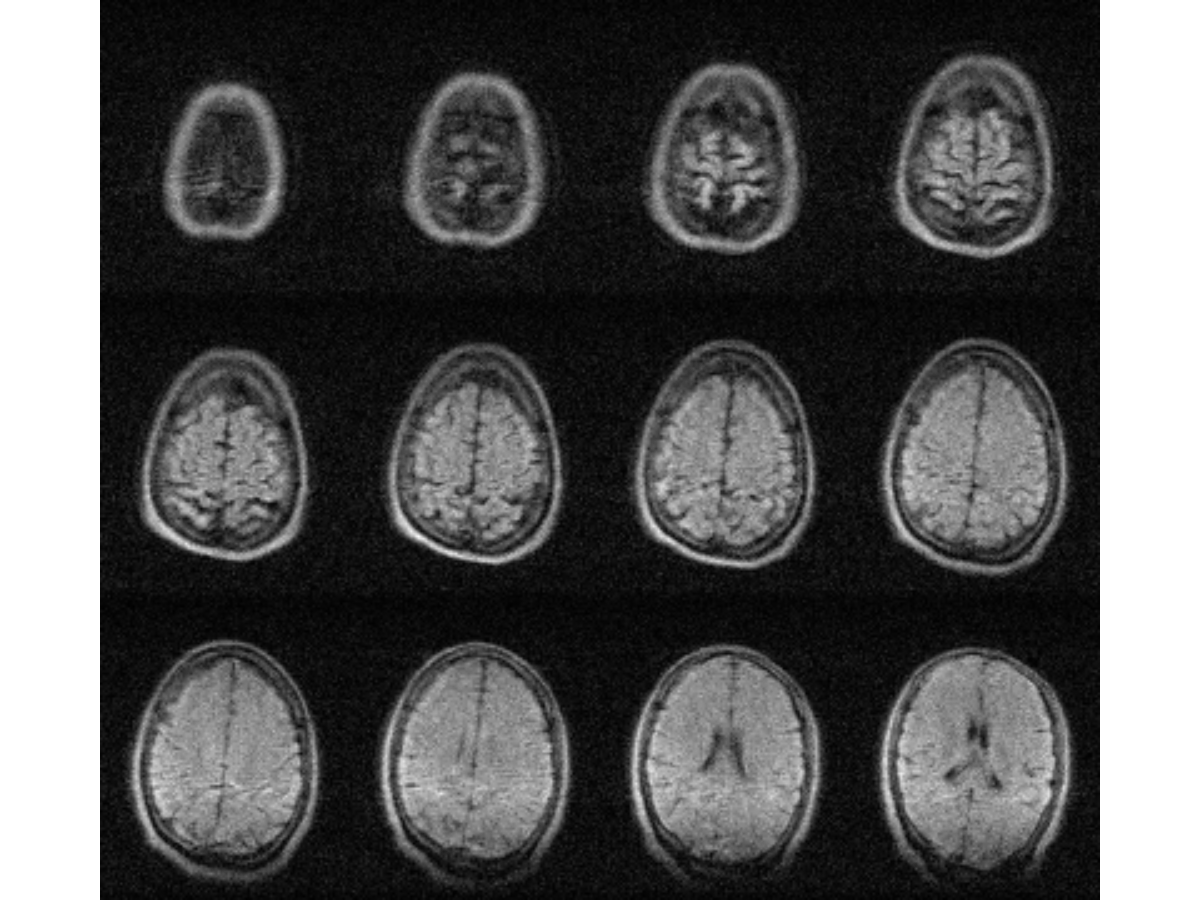}
    \caption{3D RARE images of a volunteer head. Twelve out of thirty slices are displayed. Images were reconstructed using an inverse Fast Fourier Transform without k-space filtering or additional post-processing for system imperfections to showcase the raw performance of the \gls{rf} coil.}
    \label{fig:in-vivo-results}
\end{figure}

\figurename\,\ref{fig:phantom-results-tugraz} shows sagittal, coronal and transversal slices of the OSII \textit{Hello World} phantom with a \num{15.9} \gls{snr} calculated as the ratio of the mean signal in the center of a slice and the standard deviation of the noise acquired in four squared \glspl{roi} at the corners of the image. While the images exhibit noticeable distortions caused by gradient-field nonlinearity, the geometric details of the phantom remain clearly visible, demonstrating the coil's robustness to noise.

\figurename\,\ref{fig:in-vivo-results} shows twelve transversal slices of the volunteer's head. Also in this case the geometrical details are clearly appreciable supporting the efficacy of the coil when used together with the FENCE shield for in-vivo noise suppression.

\section{Discussion}
The designed \gls{rf} coil is the result of an optimization procedure based on a number of numerical simulations. Each simulation analyzed a different distribution of the conductors and number of solenoid turns to optimize a defined objective. This procedure inevitably required some arbitrary decisions on the description of the coil topology, optimization objective and constraints.

The conductor distribution was analytically described with hyperbolic functions. Whilst this was not the only possible choice, it provided a practical and effective way to cover the desired spectra of the conductor distributions. Different analytical functions, able to lead to analogous topologies, are not expected to introduce significant improvements. 

The optimization objective (see \eqref{eq:optObjective}) was a linear combination of two figures of merit; one reflecting the magnetic field homogeneity and the other the coil sensitivity and \gls{snr}. Whereas it is generally recognized that these two parameters are of paramount importance for an \gls{rf} coil designed to operate in both transmission and reception\cite{bib:gruber1}, their relative weight in defining the objective deserves discussion, since different choices leads to different optimal results. In \eqref{eq:optObjective} the \gls{rf} coil sensitivity ($B$ term in the equation) weighted \SI{50}{\percent} more than the magnetic field homogeneity ($A$ term in the equation). The choice was a consequence of the lower \gls{snr} achievable in low-field scanners compared to medium- and high-field scanners, and the consequent importance of minimizing coil losses\cite{bib:giovannetti3}. The requirement of having easily replaceable circuitry on the \gls{rf} coil prototype introduced further losses due to additional soldering, pin connections and PCB traces with thickness lower than the skin depth. The \num{1.5} factor multiplying $B$ in \eqref{eq:optObjective} was intended to help in compensating these non-idealities. Despite this, the optimum \gls{rf} coil configuration would not have changed even if the weight factor had been removed from the objective. 

\figurename\,\ref{fig:optResults} suggests that better results could have been obtained if the maximum number of solenoid turns had not been limited to \num{14}. Whilst this requires further analysis, an increase on the number of turns would improve the \gls{rf} coil sensitivity and Q factor. However, the measured Q factor is already in-line or higher than that of similar \gls{rf} coils and an excessively high value also comes with disadvantages\cite{bib:reilly1,bib:webb1}. 

An increase of the Q factor necessarily entails a decrease of the coil bandwidth. In receive mode the coil bandwidth may become less than the acquisition bandwidth, producing banding in the MR image\cite{bib:webb1}. In literature there are solutions to deal with high Q coil banding artifacts\cite{bib:hrovat1,bib:raad1} and a preamplifier that is noise-matched rather than power-matched also proves to be effective\cite{bib:webb1}. In transmit mode, a too low bandwidth may result in non-uniform excitations during 3D acquisitions or with slices far from the isocenter. A possible solution is to add an extra resistor that is shorted during reception\cite{bib:webb1}. However, this strategy requires a dedicated circuit to short the resistor and different matching circuits in transmission and reception. 
Under some simplifying assumptions, the maximum bandwidth depends on the 
inverse of the natural logarithm of $1/\Gamma$, being $\Gamma$ the \gls{rf} coil input reflection coefficient\cite{bib:bode1, bib:gonzales1}. Therefore, as long as the \gls{rf} power amplifier and \gls{lna} remain stable and the noise figure is not overly affected by a reflection coefficient different from the optimum one, a more practical and simpler way to deal with the problem may be to accept a larger \gls{rf} coil input reflection coefficient.

The designed \gls{rf} coil was limited to one segmentation only. Further segmentation of the coil would require even higher capacitance values and would introduce additional conductor losses which, from simulations, already represent the \SI{47}{\percent} of the total losses. Of these losses, about \SI{10}{\percent} are due to the capacitors despite of high-Q capacitors were used in the design as also suggested in literature\cite{bib:giovannetti4}.

Benchtop measurements revealed the impact that the shield diameter and thickness has on the \gls{rf} coil performance. Measurements also suggested possible criticality in TU Graz and PTB rebuilds which are worth further investigation. The \gls{rf} coil housings have been left in a water-bath for several hours to remove the \gls{pva} water-soluble support material. Measurements carried out on the INRiM coil using a thermal-imaging camera showed that, when the coil was energized, most of the losses occurred inside the plastic housing. This revealed that, due to imperfections of the \gls{fff} 3D printing, some water remained trapped inside the plastic lattice and required to heat up the \gls{rf} coil for about \qty{12}{\hour} at \qty{45}{\degree} to make the water evaporate. The same procedure can be applied on the other coil rebuilds to check whether it will impact their Q factor as well. 

From an imaging perspective, the coil was successfully deployed in both phantom and in-vivo experiments. With reference to in-vivo experiments, the coil demonstrated a good noise rejection when combined to the FENCE shield\cite{bib:Pfitzer2026} making superfluous the use of conductive blankets or electrodes for subject grounding and/or expensive and voluminous Faraday cages\cite{bib:oreilly1, bib:lena1}. Regarding noise, specific \gls{emi} coupling measurements suggested that the \gls{rf} coil could benefit from being wound in a double layer configuration. This compensates the intrinsic head asymmetry along the main coil axis increasing its common noise rejection ratio. Focusing on the ratio of the \gls{rms} noise with respect to the \SI{50}{\ohm} baseline, the double layer coil performed \SI{33}{\percent} better than the single layer one. However, an \gls{rms} noise higher than \num{14} times that of the \SI{50}{\ohm} baseline proved to be too high to perform imaging and  would still require the use of the FENCE shield. Nevertheless, the rationale behind the double layer configuration warrants further investigation. Increasing the number of coil turns wound in helical pattern rather than in parallel could be beneficial. In addition, a differential receive circuit instead of a single-ended one is expected to improve the common noise rejection ratio as well.

From a safety standpoint, \gls{em} simulations with a digital human model showed that the \gls{rf} coil is safe for the imaging conditions listed in \tablename\,\ref{tab:mri_params_invivo}. However, when all the power available from the \gls{rf} power amplifier\cite{bib:osii_rfpa} is exploited, the maximum \SI{10}{\gram} averaged local \gls{sar} exceeds the first level controlled operation mode safety limit set by the IEC 60601:33 standard\cite{bib:IEC60601-2-33-2022}. Accounting for the simulated \gls{rf} coil sensitivity and the same experiment conditions summarized in \tablename\,\ref{tab:mri_params_invivo}, only about \SI{1.8}{\watt} peak forward power and \SI{10}{\milli\watt} average power is transmitted during the experiments. The substantial headroom relative to the \SI{1}{\kilo\watt} peak power of the current \gls{rf} power amplifier design suggests that the amplifier could be designed with a significantly lower \gls{rf} peak power. This would not only reduce cost and design complexity by lowering power dissipation, but also improve imaging safety.

The presented \gls{rf} coil connector proved that it is possible to design and realize open-source low-cost solutions based on off-the-shelf contacts and rapid prototyping technology. While this cuts the cost with respect to available commercial connectors\cite{bib:odu} it also provides the flexibility for implementing open standard solutions such as \gls{mrcods} for coil identification\cite{bib:mrcods}. Future activities related to the connector will deal with its integration in the low-field scanner and subsequent use for actual experiments. This will require modifying the scanner to integrate the connector ensuring a robust ground connection  which is key for effective noise suppression\cite{bib:Pfitzer2026, bib:lena1}.

Additional future activities will also exploit the modularity of the coil PCB to include new features like Q-spoiling and coil detuning. Furthermore, the same design approach will be applied to the optimization of \gls{rf} coils targeting different imaging regions, such as the extremities.

\section{Conclusion}
The paper presented the design and construction of a solenoid \gls{rf} coil optimized for head imaging. Three different \gls{rf} coil rebuilds were compared through benchtop measurements showing the effect of the different experimental setup and suggesting the importance of reference measurement results for benchmarking potential criticalities in the rebuilds. 

The \gls{rf} coil proved to be successful for in phantom and in-vivo imaging applications showing a full compatibility with the FENCE shield for noise suppression. 

In addition the paper also demonstrated the design and realization of an open-source \gls{rf} coil connector based on off-the-shelf contacts and rapid prototyping technology. The connector\cite{bib:osiiConnector} was designed to provide the same functionality, at a fraction of the cost, of those typically found on mainstream commercial \gls{mri} system and is part of the Open Connectors repository\cite{bib:openConnectors}

The \gls{rf} coil design files and documentation are available in a repository\cite{bib:zanovello2} included within the Open Source Imaging Initiative GitLab project\cite{bib:osii_gitlab}. 

\section*{Acknowledgments}
This work was supported by the project 22HLT02 A4IM which has received funding from the European Partnership on Metrology, co-financed from the European Union's Horizon Europe Research and Innovation Programme and by the Participating States.

\printbibliography

\end{document}